\documentclass[aps,prd,twocolumn,showpacs]{revtex4-1}
\usepackage{amssymb}
\usepackage{amsmath}
\usepackage{wallpaper}
\usepackage{bm}
\usepackage{multirow}
\usepackage[colorlinks,linkcolor=blue, urlcolor=blue, anchorcolor=blue, citecolor=blue]{hyperref}

\usepackage{graphicx}
\usepackage{booktabs}
\usepackage{siunitx}

\begin{document}

\title{Elastic properties of nuclear pasta in neutron-star crusts}
\author{Cheng-Jun~Xia$^{1,2}$}
\email{cjxia@nit.zju.edu.cn}
\author{Toshiki Maruyama$^{2}$}
\email{maruyama.toshiki@jaea.go.jp}
\author{Nobutoshi Yasutake$^{3, 2}$}
\email{nobutoshi.yasutake@it-chiba.ac.jp}
\author{Toshitaka Tatsumi$^{4}$}
\email{tatsumitoshitaka@gmail.com}
\author{Ying-Xun Zhang$^{5, 6}$}
\email{zhyx@ciae.ac.cn}

\affiliation{$^{1}${Center for Gravitation and Cosmology, College of Physical Science and Technology, Yangzhou University, Yangzhou 225009, China}
\\$^{2}${Advanced Science Research Center, Japan Atomic Energy Agency, Shirakata 2-4, Tokai, Ibaraki 319-1195, Japan}
\\$^{3}${Department of Physics, Chiba Institute of Technology (CIT), 2-1-1 Shibazono, Narashino, Chiba 275-0023, Japan}
\\$^{4}${KItashirakawa Kamiikeda-Cho 52-4, Kyoto 606-8287, Japan}
\\$^{5}${China Institute of Atomic Energy, Beijing 102413, People's Republic of China}
\\$^{6}${Guangxi Key Laboratory Breeding Base of Nuclear Physics and Technology, Guilin 541004, China}}

\date{\today}

\begin{abstract}
Based on the relativistic mean field (RMF) model with Thomas-Fermi approximation, we investigate the elastic properties of neutron star matter. The elastic constants are estimated by introducing deformations on the nuclear pasta structures in $\beta$-equilibrium, where various crystalline configurations are considered in a fully three-dimensional geometry without the Wigner-Seitz approximation. Two scenarios with different symmetry energy slope ($L = 41.34$ and 89.39 MeV) are examined, where the the elastic constants can vary by ten times. By fitting to the numerical results, we improve the analytic formulae for the elastic properties of nuclear pasta by introducing damping factors.
\end{abstract}

\maketitle

\section{\label{sec:intro}Introduction}
According to our understanding of materials and nuclear matter on Earth, it has been long suspected that there exist complex nonuniform matter inside neutron star crusts and white dwarf cores~\cite{Baym1971_ApJ170-299, Negele1973_NPA207-298, Ravenhall1983_PRL50-2066, Hashimoto1984_PTP71-320, Williams1985_NPA435-844}. The dense materials in those astronomical objects are called astromaterials~\cite{Caplan2017_RMP89-041002}, which can be classified as hard and soft. In the interior of cold white dwarfs and outer crust of neutron stars, nuclei and highly degenerate electron gas form Coulomb crystals~\cite{Brush1966_JCP45-2102, Ogata1993_ApJ417-265, Jones1996_PRL76-4572, Potekhin2000_PRE62-8554, Medin2010_PRE81-036107, Caplan2017_RMP89-041002, Caplan2018_ApJ860-148}, i.e., hard astromaterials. At given densities, the Coulomb crystal becomes a body-centred cubic (BCC) lattice if the nuclei are of one type~\cite{Hamaguchi1997_PRE56-4671, Baiko2002_PRE66-056405, Kozhberov2018_PRE98-063205}, while more complicated lattice structures may emerge if multiple types of nuclei are included~\cite{Ogata1993_ApJ417-265, Chamel2016_PRC94-065802, Kozhberov2019_MNRAS486-4473, Fantina2020_AA633-A149}. In the inner crust of neutron stars, the density is so large that nuclei start to touch and neutrons become unbound. As density increases, nucleons rearrange themselves into nonspherical shapes, i.e., nuclear pasta as soft astromaterials. There exist typically five types of nuclear pastas~\cite{Pethick1998_PLB427-7, Oyamatsu1993_NPA561-431, Maruyama2005_PRC72-015802, Togashi2017_NPA961-78, Shen2011_ApJ197-20}, i.e., droplets/bubbles in BCC lattices, rods/tubes in honeycomb lattices, and slabs~\cite{Oyamatsu1993_NPA561-431}. Meanwhile, more complicated structures are also possible~\cite{Magierski2002_PRC65-045804, Newton2009_PRC79-055801, Kobyakov2014_PRL112-112504, Fattoyev2017_PRC95-055804}, e.g., the gyroid and double-diamond morphologies~\cite{Nakazato2009_PRL103-132501, Schuetrumpf2015_PRC91-025801}, P-surface configurations~\cite{Schuetrumpf2013_PRC87-055805, Schuetrumpf2019_PRC100-045806}, nuclear waffles~\cite{Schneider2014_PRC90-055805, Sagert2016_PRC93-055801}, Parking-garage structures~\cite{Berry2016_PRC94-055801}, droplets with deformations~\cite{Kashiwaba2020_PRC101-045804}, the intermediate structures of two phases~\cite{Watanabe2003_PRC68-035806, Okamoto2012_PLB713-284}, and amorphous structures~\cite{Jones1999_PRL83-3589, Sauls2020}.

The transport and elastic properties of nuclear pasta are heavily influenced by their structures, which are essential for interpreting various neutron star observations~\cite{Chamel2008_LRR11-10, Caplan2017_RMP89-041002}. In particular, the stellar oscillations and the continuous gravitational waves emitted by rotating neutron stars are closely related to the elastic properties. The quasi-periodic oscillations (QPOs) observed after giant flares of soft gamma repeaters are usually interpreted as global oscillations of magnetars~\cite{Kouveliotou1998_Nature393-235, Hurley1999_Nature397-41}. It was shown that the oscillation spectrum is affected by the elastic and superfluid properties of nuclear pasta~\cite{Hansen1980_ApJ238-740, Schumaker1983_MNRAS203-457, McDermott1988_ApJ325-725,  Strohmayer1991_ApJ375-679, Passamonti2012_MNRAS419-638, Gabler2018_MNRAS476-4199, Sotani2012_PRL108-201101, Sotani2017_MNRAS464-3101, Kozhberov2020_MNRAS498-5149}. A crust destruction leads to the sudden release of magnetic and elastic energy, which may be responsible for magnetar bursts~\cite{Beloborodov2014_ApJ794-L24, Beloborodov2016_ApJ833-261, Li2016_ApJ833-189}, short gamma-ray burst precursors of neutron star mergers~\cite{Tsang2012_PRL108-011102}, and pulsar glitches~\cite{Ruderm1969_Nature223-597, Baym1971_AP66-816, Haskell2015_IJMPD24-1530008, Akbal2018_MNRAS473-621, Guegercinoglu2019_MNRAS488-2275, Layek2020_MNRAS499-455}. Due to the elastic stresses of astromaterials, there could be mountains on neutron stars, leading to the asymmetric matter distributions with nonzero ellipticities $\epsilon\equiv |I_{xx}-I_{yy}|/I_{zz}$ with $I_{xx}$, $I_{yy}$, and $I_{zz}$ being the principal moments of inertia. The dimensionless equatorial ellipticity $\epsilon$ is connected to the mass quadrupole with $Q_{22}=I_{zz}\epsilon \sqrt{15/8\pi}$, where the upper bound is set by the breaking strain $\varepsilon_\mathrm{max}$ of neutron star crust with $\varepsilon_\mathrm{max}\approx 0.1$ for nuclear pastas~\cite{Horowitz2009_PRL102-191102, Chugunov2010_MNRAS407-L54, Horowitz2010_PRD81-103001, Caplan2018_PRL121-132701, Baiko2018_MNRAS480-5511, Kozhberov2020_MNRAS498-5149}. In such cases, the maximum ellipticity of neutron stars can reach $\epsilon\approx 10^{-5}$~\cite{Ushomirsky2000_MNRAS319-902} so that continuous gravitational waves are emitted for fast rotating neutron stars~\cite{Abbott2020_ApJ902-L21}.

The elastic properties of hard astromaterials were investigated extensively in previous works~\cite{Fuchs1936_PRSA157-444, Ogata1990_PRA42-4867, Strohmayer1991_ApJ375-679}. By considering the cases where point nuclei are embedded in a uniform electron background, the elastic constants as {defined below later on in} Eq.~(\ref{eq:elastic}) of BCC and FCC crystals at vanishing temperatures were estimated with~\cite{Ogata1990_PRA42-4867}
\begin{equation}
 \left\{\begin{array}{l}
   \mathrm{BCC:}\ c_{11}-c_{12}=0.04908 \mu_{0}, \ c_{44}=0.1827 \mu_{0}; \\
   \mathrm{FCC:}\ c_{11}-c_{12}=0.04132 \mu_{0}, \ c_{44}=0.1852 \mu_{0}. \\
 \end{array}\right. \label{Eq:El_MC}
\end{equation}
Here $\mu_{0}\equiv \alpha n_d Z^2/{R_\mathrm{W}}$ with $n_d$ being the nuclei density, $Z$ the proton number of each nucleus, $R_\mathrm{W} = \left(4\pi n_d/3\right)^{-1/3}$ the Wigner-Seitz (WS) radius, and $\alpha=1/137$ the fine-structure constant.

A useful formula of the effective shear modulus can then be obtained for a BCC lattice~\cite{Ogata1990_PRA42-4867}, i.e.,
\begin{equation}
  \mu_\mathrm{eff}^\mathrm{V} = 0.1194 n_d\frac{e^2 Z^2}{R_\mathrm{W}}. \label{eq:mueff_BCC}
\end{equation}
Due to the zero-point ion vibrations at lower temperatures, the shear modulus in Eq.~(\ref{eq:mueff_BCC}) may be reduced by up to 18\% for light nuclei~\cite{Baiko2011_MNRAS416-22}, while the modification becomes negligible for larger nuclei. Meanwhile, it was shown that the effects of electron screening in the randomly oriented polycrystalline matter also lead to a reduction on the shear modulus~\cite{Kobyakov2013_PRC87-055803, Kobyakov2015_MNRAS449-L110}. The elastic properties of multicomponent crystals were examined as well, which agrees with the results obtained from the linear mixing rule~\cite{Kozhberov2019_MNRAS486-4473}. Adopting an additional symmetry for the Coulomb part of the stress-strain tensor and neglecting the effects of electron screening, a universal formula on the effective shear modulus is recently obtained, i.e., $\mu_\mathrm{eff} \approx \sum_Z 0.12 n_Z \alpha Z^{5/3} \left(4\pi n_e/3\right)^{1/3}$ with $n_e = \sum_Z Z n_Z$ being the electron number density and $n_Z$ the number density of nuclei carrying charge $Z$, which is independent of the lattice structures or composition~\cite{Chugunov2020_MNRAS500-L17}.

For astromaterials located at the inner crusts of neutron stars, nuclei are immersed in dripped neutrons and electrons, which take complex shapes and form various lattice structures. The situation thus becomes much more complicated. For spherical droplets in various lattice structures, the elastic properties may be similar to that of hard astromaterials, e.g., the elastic properties of BCC and FCC lattices may be estimated by Eq.~(\ref{Eq:El_MC}) if we neglect the contributions of dripped neutrons and electron screening. The elastic properties of nuclear pastas in rod and slab phases were obtained in the framework of incompressible liquid-drop model~\cite{Pethick1998_PLB427-7}, which was recently extended to polycrystalline nuclear pastas~\cite{Pethick2020_PRC101-055802}, i.e.,
\begin{eqnarray}
\mathrm{Rods:}&&\ c_{11}-c_{12}=2c_{44}=2E_\mathrm{C} 10^{2.1\left(u^2-0.3\right)}, \nonumber \\
      &&\ c_{11}+c_{12}=3E_\mathrm{C}, \label{eq:El_ld_Rod} \\
\mathrm{Slabs:}&&\ c_{11}=6E_\mathrm{C}, \label{eq:El_ld_Slab}
\end{eqnarray}
with $u\equiv{R_d}/{R_\mathrm{W}}$. Note that the surface and Coulomb energy densities are connected by $E_\mathrm{S} = 2E_\mathrm{C}$ for optimized droplet size $R_d$ and WS cell size $R_\mathrm{W}$. For the bubble phase, it is also possible to estimate the elastic constants with Eq.~(\ref{Eq:El_MC}), where the proton number $Z$ is replaced by its effective values~\cite{Watanabe2003_PRC68-045801}.

Nevertheless, as one introduces deformations to nuclear pastas, the relative contents of protons and neutrons change since they convert into each other via weak reactions on a time scale much shorter than deformations. At the same time, as illustrated in Ref.~\cite{Caplan2018_PRL121-132701}, nucleons rearrange themselves in space and may cause short-range disorders, which is expected to alter the elastic properties as well. The effects of domains and their boundaries in nuclear pasta should also play important roles~\cite{Caplan2018_PRL121-132701, Pethick2020_PRC101-055802}. In such cases, based on our previous study~\cite{Okamoto2012_PLB713-284, Okamoto2013_PRC88-025801, Xia2021_PRC103-055812, Xia2022_PRD106-063020, Xia2023_PLB839-137769}, in this work we investigate the elastic properties of nuclear pasta in a fully three-dimensional geometry. The $\beta$-stability condition is always fulfilled in our calculation, while the dripped neutrons, the electron charge screening effect, as well as the complicated nuclear shapes are considered self-consistently.

Based on our previous findings of nuclear pasta with two different slopes of symmetry energy~\cite{Xia2021_PRC103-055812, Xia2023_PLB839-137769}, we thus examine their elastic properties. For spherical droplets, we have considered the transition from BCC lattice to FCC lattice through Bain deformation~\cite{Bain1924_TAIMME70-25}, where the height of the Bain barrier was obtained. Similarly, the transition between the honeycomb and simple configurations of rod and tube phases were examined. The single-crystal elastic constants for droplets in BCC and FCC lattices, rods/tubes in honeycomb lattices, and slabs are then estimated. The paper is organized as follows. In Sec.~\ref{sec:the}, we present our theoretical framework, where the relativistic mean field (RMF) model is introduced in Sec.~\ref{sec:the_RMF} and the elastic theory in Sec.~\ref{sec:the_elestic}. The obtained results on the elastic properties of nuclear pasta are presented in Sec.~\ref{sec:results}. Our conclusion is given in Sec.~\ref{sec:sum}.

\section{\label{sec:the} Theoretical framework}
\subsection{\label{sec:the_RMF} Relativistic mean field theory}
As was done in Ref.~\cite{Xia2021_PRC103-055812}, we adopt the following Lagrangian density for the RMF model~\cite{Meng2016_RDFNS}, i.e.,
\begin{eqnarray}
\mathcal{L}
 &=& \sum_{i=n, p} \bar{\psi}_i \left[ i \gamma^\mu \partial_\mu - m_N - g_{\sigma} \sigma \right. \nonumber \\
 &&\mbox{}\left. - \gamma^\mu \left( g_{\omega}  \omega_\mu + g_{\rho} \boldsymbol{\tau}_i \cdot \boldsymbol{\rho}_\mu
              + e A_\mu\frac{1-\tau_{i,3}}{2}\right)    \right] \psi_i  \nonumber \\
 &&\mbox{} + \bar{\psi}_e \left[ i \gamma^\mu \partial_\mu - m_e  + \gamma^\mu e A_\mu   \right] \psi_e \nonumber \\
 &&\mbox{} + \frac{1}{2}\partial_\mu \sigma \partial^\mu \sigma - \frac{1}{2}m_\sigma^2 \sigma^2
     - U(\sigma) - \frac{1}{4} \omega_{\mu\nu}\omega^{\mu\nu} \nonumber \\
 &&\mbox{}+ \frac{1}{2}m_\omega^2 \omega_\mu\omega^\mu
     - \frac{1}{4} \boldsymbol{\rho}_{\mu\nu}\cdot\boldsymbol{\rho}^{\mu\nu}
     + \frac{1}{2}m_\rho^2 \boldsymbol{\rho}_\mu\cdot\boldsymbol{\rho}^\mu \nonumber \\
 &&\mbox{} - \frac{1}{4} A_{\mu\nu}A^{\mu\nu}
     + \Lambda_\mathrm{v}g_\omega^2 g_\rho^2 (\omega_\mu\omega^\mu) (\boldsymbol{\rho}_\mu\cdot\boldsymbol{\rho}^\mu),
\label{eq:Lagrange}
\end{eqnarray}
with the field tensors
\begin{eqnarray}
\omega_{\mu\nu} &=& \partial_\mu \omega_\nu - \partial_\nu \omega_\mu, \\
\boldsymbol{\rho}_{\mu\nu}
  &=& \partial_\mu \boldsymbol{\rho}_\nu - \partial_\nu \boldsymbol{\rho}_\mu, \\
A_{\mu\nu} &=& \partial_\mu A_\nu - \partial_\nu A_\mu.
\end{eqnarray}
The parameters are taken as in Ref.~\cite{Maruyama2005_PRC72-015802}, where the masses of nucleons, electrons, $\sigma$, $\omega$, and $\rho$ mesons are $m_N = 938\ \rm{MeV}$, $m_e = 0.511\ \rm{MeV}$, $m_\sigma = 400\ \rm{MeV}$, $m_\omega = 783\ \rm{MeV}$, $m_\rho = 769\ \rm{MeV}$, respectively. {Note that the mass of $\sigma$ meson is slightly smaller than the typical values, which is necessary to reproduce the binding energies of finite nuclei in the framework of Thomas-Fermi approximation~\cite{Maruyama2005_PRC72-015802}. If a larger value $m_\sigma \approx 500\ \rm{MeV}$ is adopted,  finite nuclei will become overbound. For nuclear saturation properties, the exact value of $m_\sigma$ is irrelevant since it only appears in combination with the coupling as $g_\sigma^2/m_\sigma^2$ in  the thermodynamic potential.} For the isoscalar channel of the effective $N$-$N$ interactions, the $N$-$\sigma$ and $N$-$\omega$ coupling constants are $g_\sigma = 6.3935$ and $g_{\omega} = 8.7207$, while the nonlinear self-coupling term of $\sigma$-meson is given by $U(\sigma) = b m_N (g_{\sigma} \sigma)^3/3 + c (g_{\sigma} \sigma)^4/4$ with the coefficients $b= -0.008659$ and $c = -0.002421$. These parameters predict a nuclear saturation density $n_0 = 0.153\ \rm{fm}^{-3}$, the binding energy $B(n_0) = E/n_0 - m_N = -16.3$ MeV, the incompressibility $K(n_0) = 240$ MeV, and the effective nucleon mass $m_N^*(n_0) = 0.78 m_N$. For the isovector channel, we have adopted an $\omega$-$\rho$ cross coupling term~\cite{Shen2020_ApJ891-148} with two parameter sets, i.e., Set 0 ($g_{\rho} = 4.2696$; $\Lambda_\mathrm{v} = 0$)~\cite{Maruyama2005_PRC72-015802} and Set 1 ($g_{\rho} = 5.55048$; $\Lambda_\mathrm{v} = 0.34$)~\cite{Xia2021_PRC103-055812}, which predicts a symmetry energy $S(n_0) = 32.46$ (31.85) MeV  with its slope $L(n_0)=89.39$ (41.34) MeV  for Set 0 (1).

Carrying out a variational procedure and adopting the mean-field and no-sea approximations, the Dirac equations for fermions and Klein-Gordon equation for bosons can be obtained with Eq.~(\ref{eq:Lagrange}). The mean expectation values of bosons are then fixed by solving
\begin{eqnarray}
(-\nabla^2 + m_\sigma^2) \sigma &=& - g_{\sigma} n_\mathrm{s} - U'(\sigma), \label{eq:KG_sigma} \\
(-\nabla^2 + m_\omega^2) \omega_0 &=& g_{\omega} n_\mathrm{b} - 2 \Lambda_\mathrm{v}g_\omega^2 g_\rho^2 \omega_0 \rho_{0,3}^2, \label{eq:KG_omega}\\
(-\nabla^2 + m_\rho^2) \rho_{0,3} &=&  \sum_{i=n,p} g_{\rho}\tau_{i, 3} n_i - 2 \Lambda_\mathrm{v}g_\omega^2 g_\rho^2 \omega_0^2\rho_{0,3}, \label{eq:KG_rho}\\
- \nabla^2 A_0 &=& e n_p - e n_e. \label{eq:KG_photon}
\end{eqnarray}
Adopting Thomas-Fermi and no-sea approximations, the local scalar and vector densities of fermions at zero temperature are obtained with
\begin{eqnarray}
n_{s} &=& \sum_{i=n,p} \langle \bar{\psi}_i \psi_i \rangle = \sum_{i=n,p} \frac{{m^*_N}^3}{2\pi^2} g\left(\frac{\nu_i}{m^*_N}\right),\\
n_i &=& \langle \bar{\psi}_i\gamma^0 \psi_i \rangle = \frac{\nu_i^3}{3\pi^2},
\end{eqnarray}
where $g(x) = x \sqrt{x^2+1} - \mathrm{arcsh}(x)$. The baryon number density is then fixed by $n_\mathrm{b}=n_n + n_p$. Meanwhile, the local chemical potential for nucleons and electrons are obtained with
\begin{eqnarray}
\mu_i(\vec{r}) &=& g_{\omega} \omega_0(\vec{r}) + g_{\rho}\tau_{i, 3} \rho_{0, 3}(\vec{r}) + q_i  A_0(\vec{r}) \nonumber \\
&&{}  + \sqrt{{\nu_i(\vec{r})}^2+{m_N^*(\vec{r})}^2}  = \rm{constant}, \label{eq:chem_consN} \\
\mu_e(\vec{r}) &=& \sqrt{{\nu_e(\vec{r})}^2+{m_e(\vec{r})}^2} - e  A_0(\vec{r})  = \rm{constant}, \label{eq:chem_conse}
\end{eqnarray}
where $\nu_i$ is the Fermi momentum corresponding to the top of Fermi-sea. The effective nucleon mass is obtained with $m_N^*\equiv m_N + g_{\sigma} \sigma$.

Finally the energy density of the nuclear pasta is determined by
\begin{equation}
E =\left.\int \langle {\cal{T}}_{00} \rangle \mbox{d}^3r\right/\int  \mbox{d}^3r, \label{eq:energy}
\end{equation}
with the energy momentum tensor
\begin{eqnarray}
\langle {\cal{T}}_{00} \rangle
&=&  \sum_{i=n,p} \frac {{m^*_N}^4}{8\pi^{2}}f\left(\frac{\nu_i}{m^*_N}\right) + \frac {m_e^4}{8\pi^{2}}f\left(\frac{\nu_e}{m_e}\right)\nonumber \\
&&{}   + \frac{1}{2}(\nabla \sigma)^2 + \frac{1}{2}m_\sigma^2 \sigma^2 + U(\sigma)
       + \frac{1}{2}(\nabla \omega_0)^2  \nonumber \\
&&{}   + \frac{1}{2}m_\omega^2 \omega_0^2
       + \frac{1}{2}(\nabla \rho_{0,3})^2 + \frac{1}{2}m_\rho^2 \rho_{0,3}^2 \nonumber \\
&&{}   + 3 \Lambda_\mathrm{v}g_\omega^2 g_\rho^2 \omega_0^2 \rho_{0,3}^2
       + \frac{1}{2}(\nabla A_0)^2,
\label{eq:ener_dens}
\end{eqnarray}
where $f(x) = \left[x(2x^2+1)\sqrt{x^2+1}-\mathrm{arcsh}(x) \right]$. The Coulomb energy density adopted in Eqs.~(\ref{eq:El_ld_Rod}) and (\ref{eq:El_ld_Slab}) is determined with
\begin{equation}
E_\mathrm{C} =\frac{1}{2}\left.\int (\nabla A_0)^2 \mbox{d}^3r\right/\int  \mbox{d}^3r. \label{eq:E_C}
\end{equation}
The pressure is then obtained with $P =\sum_{i=n,p,e} \mu_i n_i -E$.

To investigate the structures of nuclear pastas in the inner crusts of neutron stars, at fixed baryon number density $n_\mathrm{b}$ and proton fraction $Y_p$, we solve Eqs.~(\ref{eq:KG_sigma}-\ref{eq:KG_photon}), (\ref{eq:chem_consN}), and (\ref{eq:chem_conse}) iteratively in a 3D periodic cell with discretized space coordinates~\cite{Okamoto2012_PLB713-284, Okamoto2013_PRC88-025801}. The grid distances on $x$-, $y$-, and $z$-axis are taken as $\Delta x$, $\Delta y$, and $\Delta z$ with the numbers of grids being $N_x$, $N_y$, and $N_z$, which corresponds to the cell with a volume of $V=\int  \mbox{d}^3r = \Delta x N_x \Delta y N_y \Delta z N_z$. Note that the Klein-Gordon equations~(\ref{eq:KG_sigma}-\ref{eq:KG_photon}) are solved via fast cosine/fourier transformations at fixed density profiles, while the constancy of chemical potentials in Eqs.~(\ref{eq:chem_consN}) and (\ref{eq:chem_conse}) are fulfilled by readjusting the density distributions of fermions using the imaginary time step method~\cite{Levit1984_PLB139-147}. To reach the ground state, the proton fraction $Y_p$ are readjusted to meet the requirement of $\beta$ stability condition $\mu_n(\vec{r}) = \mu_p(\vec{r}) + \mu_e(\vec{r})$, while the global charge neutrality condition $\int (n_p - n_e) \mbox{d}^3r = 0$ is always satisfied in the meantime.

\subsection{\label{sec:the_elestic} Elasticity theory}
Based on elasticity theory, the variation of energy density due to the deformation of nuclear pasta can be expanded as~\cite{Wallace1967_PR162-776}
\begin{equation}
\delta E = \sum_{i,j} \sigma_{ij}u_{ij} + \sum_{i,j,k,l}\frac{1}{2}S_{ijkl}u_{ij}u_{kl}, \label{eq:dE0}
\end{equation}
where the indices $i$, $j$, $k$, $l$ represent the Cartesian components ($x$, $y$, and $z$), $\sigma_{ij}$ ($=-P\delta_{ij}$) the stress tensor in an undeformed solid, and $S_{ijkl}$ the elastic modulus tensor. The deformation is introduced via a displacement gradient $u_{ij}$, where a droplet at position $\vec{r}$ is moved to a new position $\vec{r}'$ with
\begin{equation}
  r'_i = r_i + \sum_j u_{ij} r_j.
\end{equation}
In this work we consider a uniform deformation with constant $u_{ij}$. If we assume nuclear pasta deforms at a fixed volume, i.e., $\mathrm{det}\left[u+I_{3\times3}\right]=1$, in {Eq.~(\ref{eq:dE0})} the first term corresponding to $\sigma_{ij}$ vanishes. Carrying out a transformation ($ij$, $kl$) $\rightarrow$ ($m$, $n$) of the subscripts ($xx$, $yy$, $zz$, $xy$, $yz$, $zx$)  $\rightarrow$ (1, 2, 3, 4, 5, 6), the elastic constants $c_{mn}$ are then connected to the elastic modulus tensor with~\cite{Ogata1990_PRA42-4867}
\begin{equation}
c_{mn} = S_{ijkl}. \label{eq:elastic}
\end{equation}
Then the variation of energy density becomes
\begin{equation}
\delta E = \frac{1}{2} \sum_{n=1}^3\sum_{m=1}^3 c_{mn} u_m u_n + 2\sum_{m=4}^6 c_{mm} u_m^2, \label{eq:dE1}
\end{equation}
where $c_{mn}=c_{nm}$ and $u_m = (u_{ij}+u_{ji})/2$. For nuclear pasta with cubic symmetry such as BCC and FCC lattices, we have $c_{11}=c_{22}=c_{33}$, $c_{12}=c_{21}=c_{13}=c_{31}=c_{23}=c_{32}$, and $c_{44}=c_{55}=c_{66}$. For the rods and tubes aligned with the $z$-axis, the nonzero terms are $c_{11}$ ($=c_{22}$), $c_{12}$, and $c_{44}$ ($=c_{11}/2-c_{12}/2$ for honeycomb configuration)~\cite{Pethick2020_PRC101-055802}, while for slabs perpendicular to the $x$-axis the nonzero term is $c_{11}$.

\section{\label{sec:results} Results and discussions}
To investigate the elastic properties of nuclear pastas, we perform volume-preserving deformations similar as in Refs.~\cite{Ogata1990_PRA42-4867, Caplan2018_PRL121-132701} along axis $i$ ($=x$, $y$, $z$), i.e.,
\begin{equation}
D_1:\ \ \ u_{ii} = \left( 1-\frac{\varepsilon}{2} \right) ^{-2}-1, \ \  \left.u_{jj}\right|_{j\neq i} = -\frac{\varepsilon}{2}. \label{eq:def_drop}
\end{equation}
In practice, we start from the nuclear pasta structures obtained with initial grid distances $\Delta x =\Delta x_0$, $\Delta y = \Delta y_0$, and $\Delta z = \Delta z_0$ as in our previous calculations~\cite{Xia2021_PRC103-055812}, then adjust them sequentially according to Eq.~(\ref{eq:def_drop}), i.e., $\Delta x = u_{xx}\Delta x_0$, $\Delta y = u_{yy}\Delta y_0$, and $\Delta z = u_{zz}\Delta z_0$. Note that Eqs.~(\ref{eq:KG_sigma}-\ref{eq:KG_photon}), (\ref{eq:chem_consN}), and (\ref{eq:chem_conse}) are solved in the 3D periodic cell once $\varepsilon$ is changed. By introducing deformations to nuclear pasta, the mean fields and the energy density in Eq.~(\ref{eq:energy}) are altered as well.

\begin{figure}
\includegraphics[width=\linewidth]{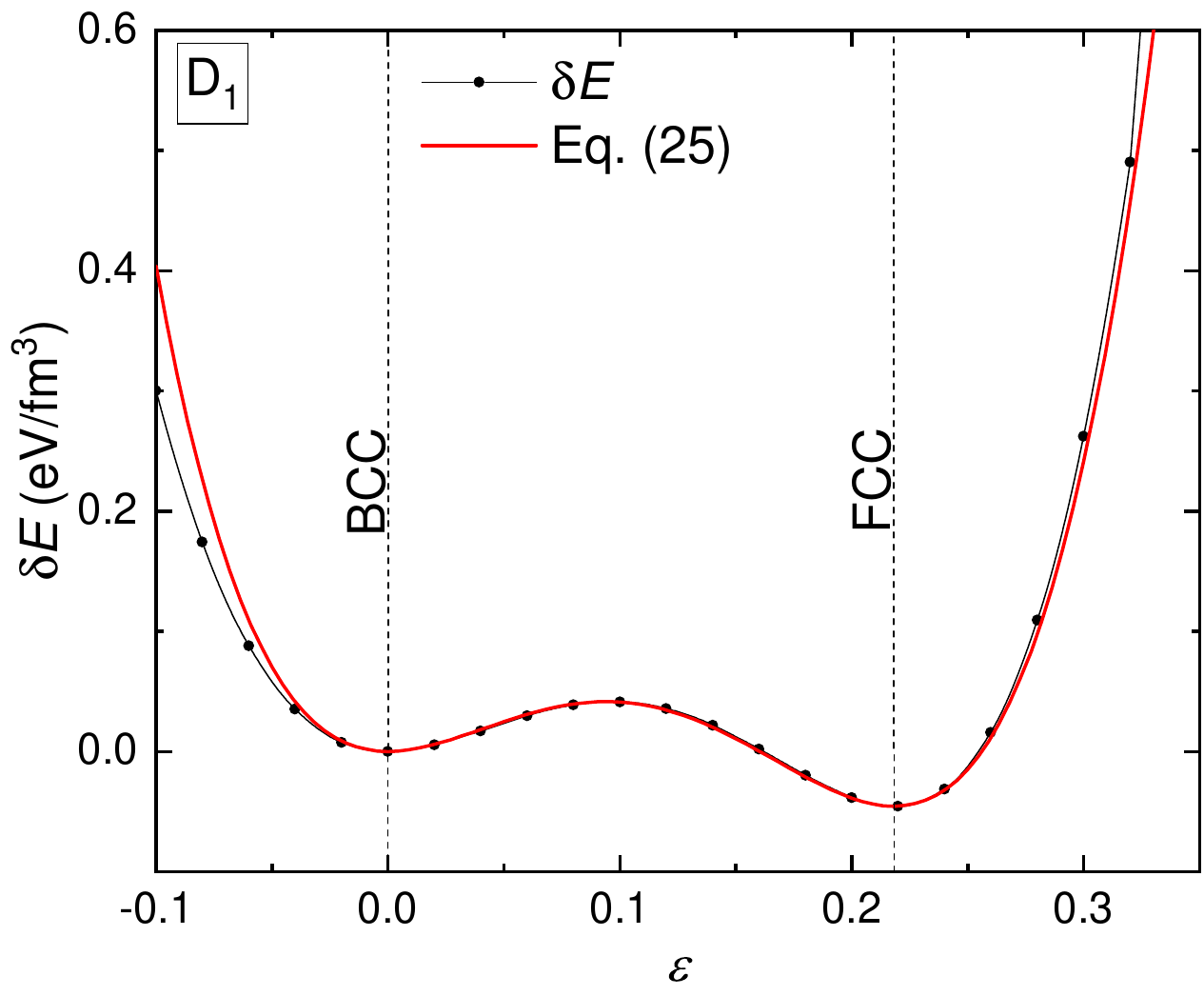}
\caption{\label{Fig:DEBtoF} Variation of energy density as a function of $\varepsilon$, where the droplets in BCC lattice ($\varepsilon=0$) evolve into FCC lattice ($\varepsilon=\varepsilon_\mathrm{F}$) at a fixed baryon number density $n_\mathrm{b} = 0.06\ \rm{fm}^{-3}$. The red solid curve corresponds to the fitting formula in Eq.~(\ref{eq:dEBtoF}). The parameter Set 0 is adopted for the isovector channel of effective $N$-$N$ interactions, while the $\beta$-stability condition is always fulfilled.}
\end{figure}

The BCC lattice ($a=b=c$) can evolve into FCC lattice ($\sqrt{2}a=\sqrt{2}b=c$) by applying the deformation Eq.~(\ref{eq:def_drop}) on $z$-axis with $\varepsilon=\varepsilon_\mathrm{F}=2-2^{5/6}\approx 0.2182$, i.e., the Bain path~\cite{Bain1924_TAIMME70-25}. In principle, since the BCC and FCC lattices have cubic symmetry, the deformation can be carried out on any axis. As an example, in Fig.~\ref{Fig:DEBtoF} we present the variation of energy density with respect to droplets in BCC lattice. The energy density can then be well reproduced by the following formula:
\begin{equation}
\delta E = C \left[ \frac{1}{4} \varepsilon^4 - \frac{1}{3} \left( \varepsilon_\mathrm{F} + \varepsilon_\mathrm{B} \right) \varepsilon^3+ \frac{1}{2} \varepsilon_\mathrm{F} \varepsilon_\mathrm{B} \varepsilon^2 \right], \label{eq:dEBtoF}
\end{equation}
where two parameters are introduced, i.e., $\varepsilon_\mathrm{B}$ corresponding to the position of Bain barrier and $C$ the strength of variation. We then fix the parameters by the energy densities at $\varepsilon = 0.1$ and $\varepsilon_\mathrm{F}$, which gives
\begin{eqnarray}
C &=&  28629\delta E (0.1)-12527.5\delta E (\varepsilon_\mathrm{F}), \label{eq:parm_BtoF0}\\
\varepsilon_\mathrm{B} &=&  0.063 + 1319.819 \delta E (0.1)/C. \label{eq:parm_BtoF1}
\end{eqnarray}
As indicated by the solid red curve in Fig.~\ref{Fig:DEBtoF}, the formula in Eq.~(\ref{eq:dEBtoF}) well reproduces the obtained energy density. The barrier height from BCC to FCC can then be obtained with $\delta E (\varepsilon_\mathrm{B}) = C\varepsilon_\mathrm{B}^3 (2\varepsilon_\mathrm{F}-\varepsilon_\mathrm{B})/12$, while the barrier height from FCC to BCC is given by $\delta E (\varepsilon_\mathrm{B}) - \delta E (\varepsilon_\mathrm{F}) = C\left( \varepsilon_\mathrm{F}+\varepsilon_\mathrm{B} \right)  \left( \varepsilon_\mathrm{F}-\varepsilon_\mathrm{B} \right) ^{3}/12$. Adopting the cubic symmetry and neglecting higher-order terms, the elastic constants $c_{11}$ ($=c_{22}=c_{33}$) and $c_{12}$ ($=c_{21}=c_{13}=c_{31}=c_{23}=c_{32}$) for the BCC lattice is give by $c_{11} - c_{12} = 2C\varepsilon_\mathrm{F} \varepsilon_\mathrm{B}/3$, while for FCC lattice we have $c_{11} - c_{12} = C \varepsilon_\mathrm{F} \left( \varepsilon_\mathrm{F} -2 \right)^2 \left(\varepsilon_\mathrm{F}-\varepsilon_\mathrm{B}\right)/6$.

\begin{figure}
\centering
\includegraphics[width=\linewidth]{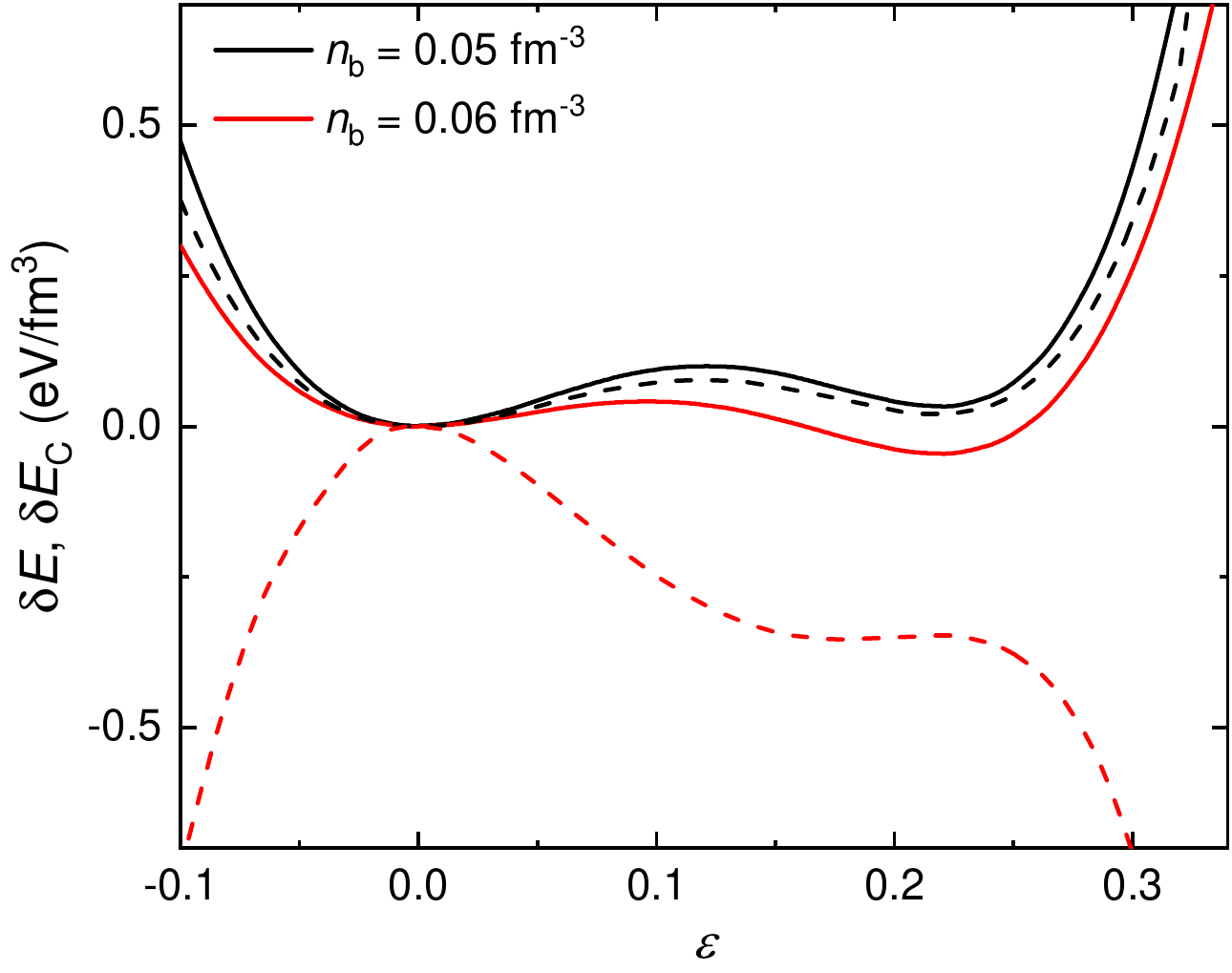}
\caption{\label{Fig:DEcomp_dens} Variations of total ($\delta E$, solid) and Coulomb ($\delta E_\mathrm{C}$, dashed) energy densities for the droplet phase in $\beta$-stability as functions of deformation, where parameter Set 0 with $L=89.39$ MeV is adopted with $n_\mathrm{b} = 0.05\ \rm{fm}^{-3}$ (black) and $0.06\ \rm{fm}^{-3}$ (red).}
\end{figure}

\begin{figure}
\centering
\includegraphics[width=\linewidth]{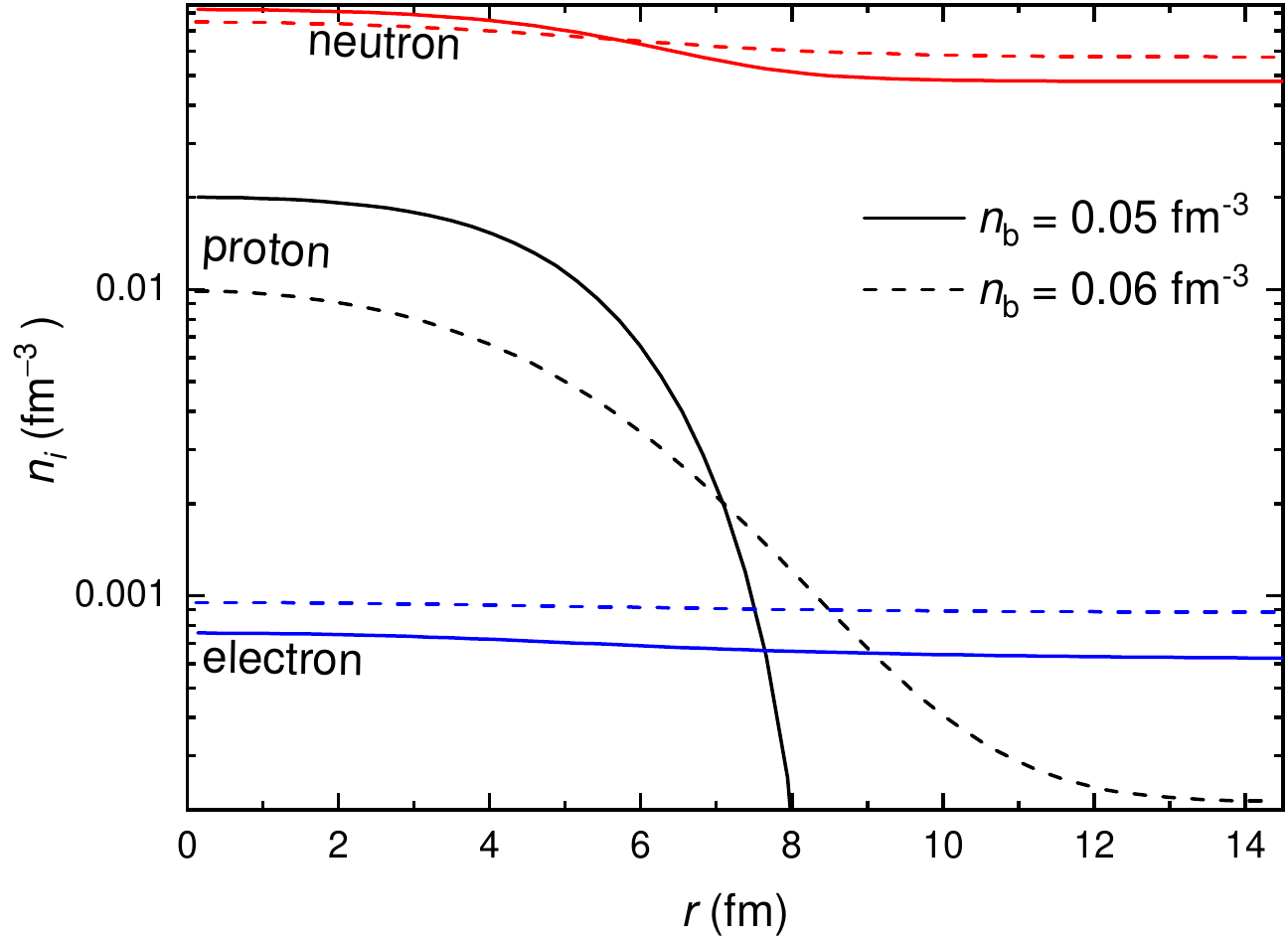}
\caption{\label{Fig:Dens} The density profiles of nucleons and electrons for the droplets indicated in Fig.~\ref{Fig:DEcomp_dens}.}
\end{figure}

To disentangle the quantitative deviations on the elastic properties, we have carried out additional calculations on nuclear pastas under deformation. Similar to Fig.~\ref{Fig:DEBtoF}, Fig.~\ref{Fig:DEcomp_dens} shows the the variations of total and Coulomb energy densities for the droplet phase in $\beta$-stability. It is evident that the variation of energy (solid) is mainly attributed to Coulomb (dashed) interactions at $n_\mathrm{b} = 0.05\ \rm{fm}^{-3}$, which confirms the validity of assuming Coulomb crystals for the droplet phase above neutron drip density. However, at larger densities ($n_\mathrm{b} = 0.06\ \rm{fm}^{-3}$), the contribution of Coulomb energy becomes negative~\cite{Xia2022_PRD106-063020}, where the analytical formulae in Eq.~(\ref{Eq:El_MC}) overestimate the elastic constants. As indicated in Fig.~\ref{Fig:Dens}, this is mainly attributed to the increment of neutron gas density, where the liquid-gas interface becomes less evident. By subtracting the background proton number density, the analytical formulae in Eq.~(\ref{Eq:El_MC}) can be improved to reproduce the trends of Fig.~\ref{Fig:DEBtoF}, where the elastic constant decreases with density. {Note that the diffused proton density at $n_\mathrm{b} = 0.06\ \rm{fm}^{-3}$ is related to the core-crust transition, where the crustal matter is about to transit into the uniform core matter at slightly larger densities ($0.06\ \rm{fm}^{-3}<n_\mathrm{b}<0.061\ \rm{fm}^{-3}$)~\cite{Maruyama2005_PRC72-015802, Xia2021_PRC103-055812}.}

\begin{figure}
\centering
\includegraphics[width=\linewidth]{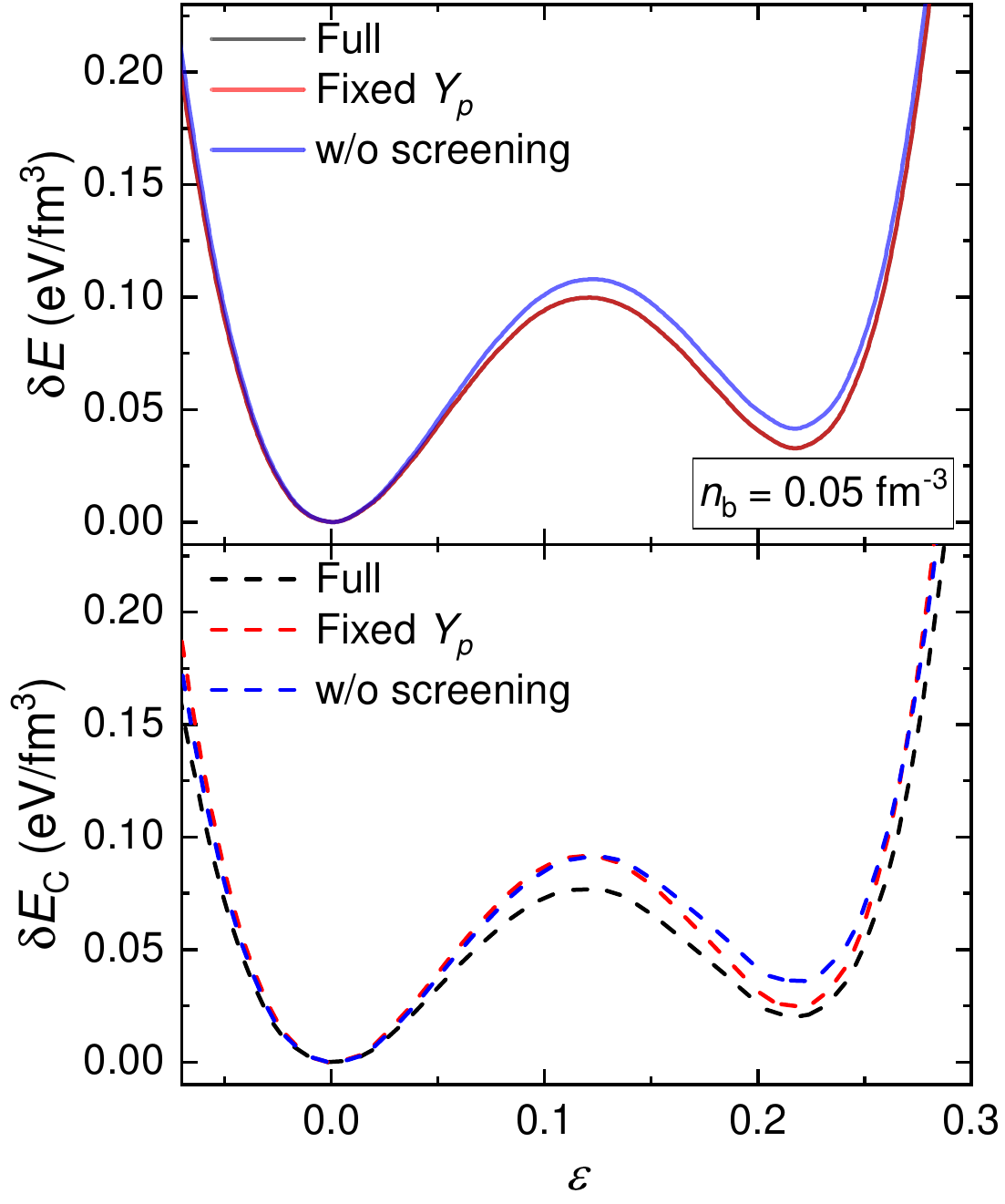}
\caption{\label{Fig:DEcomp_all} Similar as Fig.~\ref{Fig:DEcomp_dens} but for $n_\mathrm{b} = 0.05\ \rm{fm}^{-3}$, where the results obtained with the full calculation in $\beta$-stability (black), with those of fixed $Y_p$ (red), and without charge screening (blue) are compared. }
\end{figure}

Figure~\ref{Fig:DEcomp_all} shows the results obtained with fixed proton fraction $Y_p$ (red) and assuming constant electron density (blue), respectively. By comparing with the results obtained with full calculations (black), the effects of weak reactions and charge screening can be identified. We note that by imposing $\beta$-stability condition for nuclear pasta, the proton fraction $Y_p$ are altered by deformation and consequently the Coulomb energy densities are modified. Nevertheless, if we examine the total energies with and without $\beta$-stability, both of which predict similar energy density variations as functions of deformation,  where both the black and red solid curves overlap with each other. In such cases, we conclude that weak reactions have little contributions to the elastic properties of neutron star matter. The charge screening effects, however, reduces the elastic constants as indicated in Fig.~~\ref{Fig:DEcomp_all}. Additionally, slight deviations from the spherical shapes of droplets are observed, which alters the elastic constants as well~\cite{Zemlyakov2022_MNRAS518-3813}.

\begin{figure}
\includegraphics[width=\linewidth]{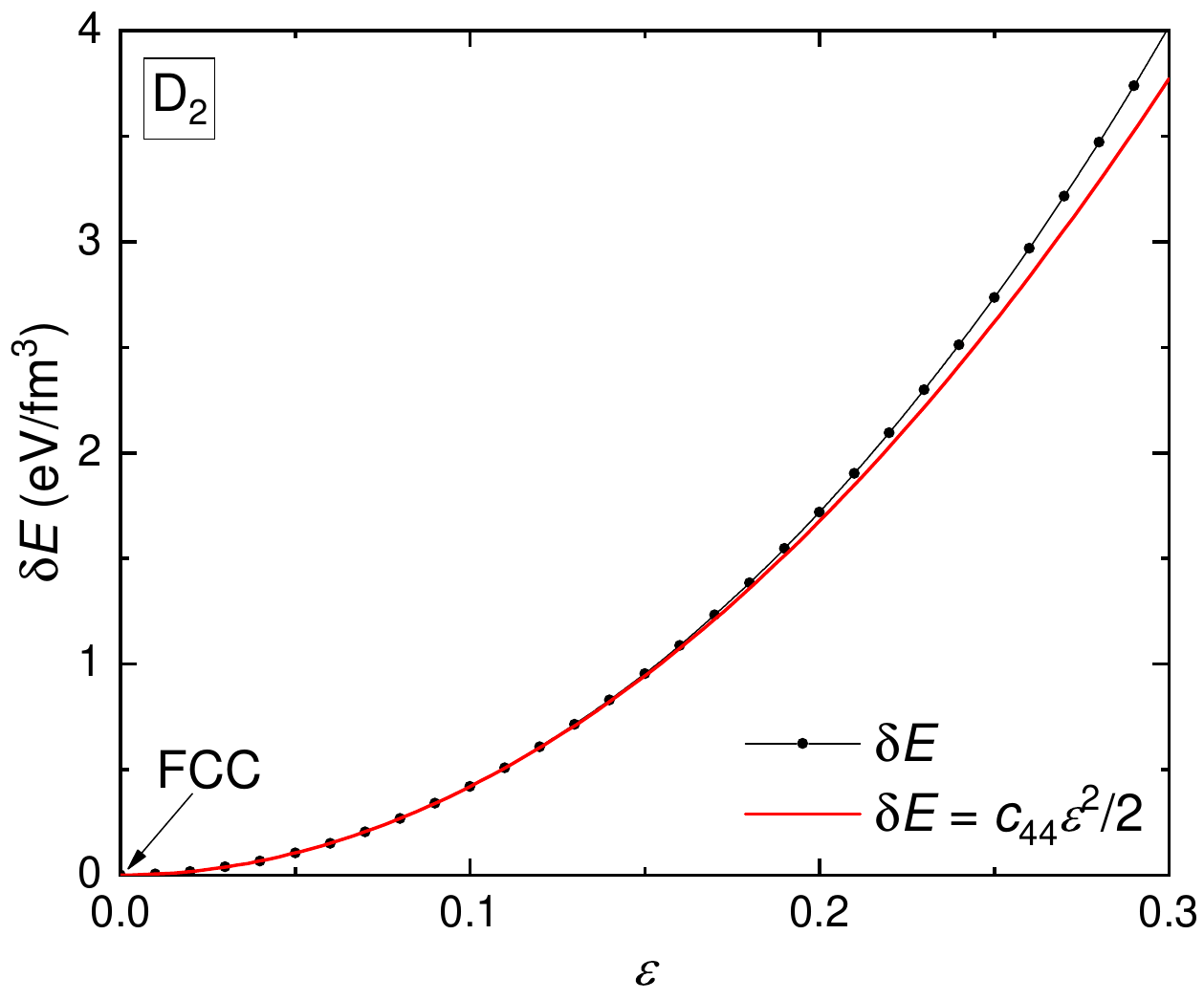}
\caption{\label{Fig:D2FCC} Same as Fig.~\ref{Fig:DEBtoF} but applying deformations on droplets in FCC lattice ($\varepsilon=0$) at fixed baryon number density $n_\mathrm{b} = 0.06\ \rm{fm}^{-3}$ using Eq.~(\ref{eq:def_drop_shear0}). The solid curve corresponds to the parabolic fitting formula~(\ref{Eq:c44_drop}).}
\end{figure}

To estimate $c_{44}$ ($=c_{55}=c_{66}$) of BCC and FCC lattices, we apply the deformation~\cite{Ogata1990_PRA42-4867}
\begin{equation}
D_2:\ \ \ u_{xy} = u_{yx} = \frac{\varepsilon}{2}, \ \  u_{zz} = \frac{4}{4-\varepsilon^2}-1.  \label{eq:def_drop_shear0}
\end{equation}
In practice, to carry out the deformation $D_2$, we rotate the nuclear pasta $45^\circ$ along $z$-axis and apply the following deformation
\begin{equation}
D_3:\ \ \ u_{xx} = -u_{yy} = \frac{\varepsilon}{2}, \ \  u_{zz} = \frac{4}{4-\varepsilon^2}-1.  \label{eq:def_drop_shear}
\end{equation}
Then we have $c_{44}= 2\delta E(\varepsilon)/\varepsilon^2$ if the higher order terms $O(\varepsilon^4)$ can be neglected, where
\begin{equation}
  \delta E=c_{44}\varepsilon^2/2. \label{Eq:c44_drop}
\end{equation}
As indicated in Fig.~\ref{Fig:D2FCC}, this formula is valid at $\varepsilon\lesssim 0.18$, while slight discrepancies are observed if larger deformations are adopted.

\begin{figure}
\includegraphics[width=\linewidth]{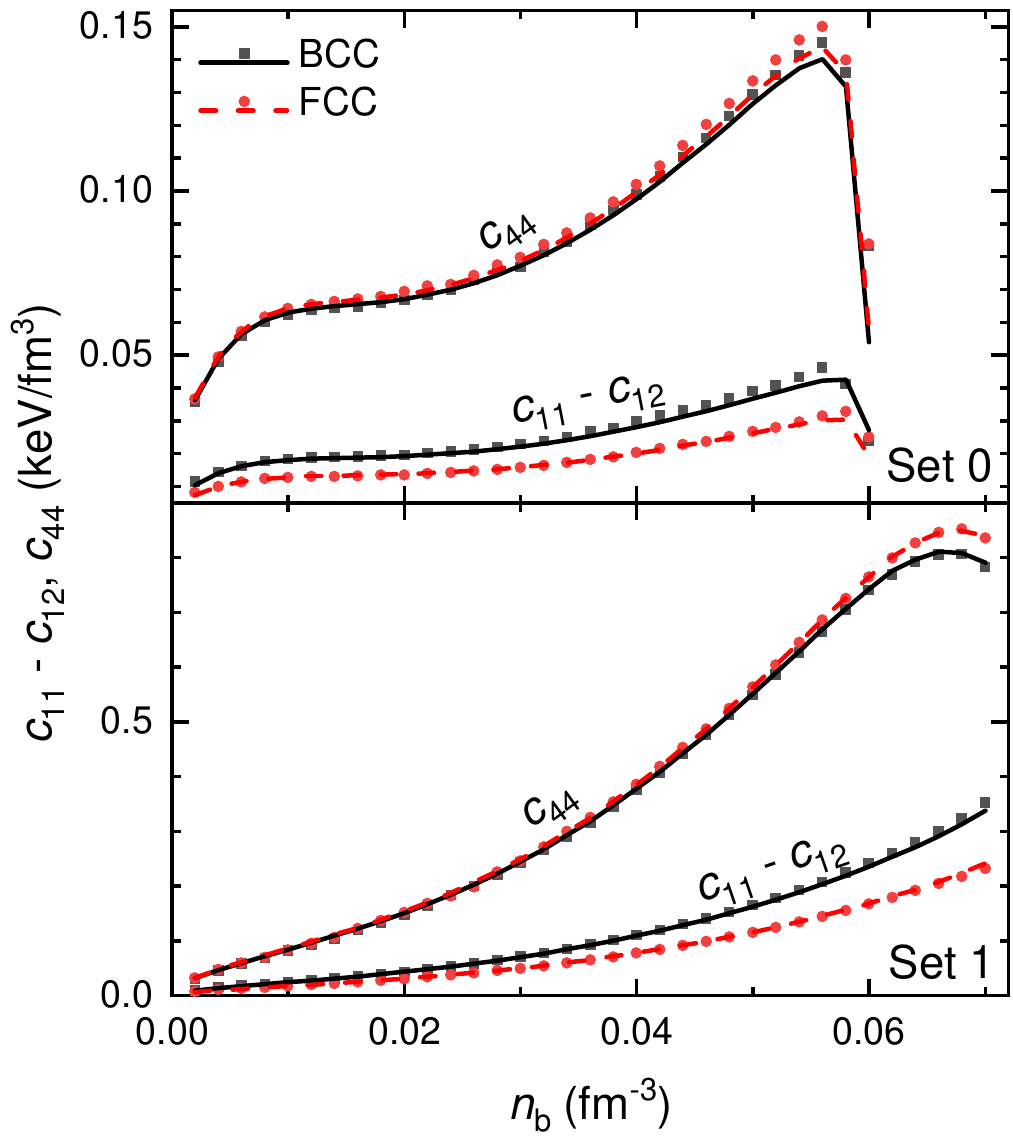}
\caption{\label{Fig:ElasDrop} Elastic constants of the droplet phase for the inner crust matter of neutron stars. The solid and dashed curves represent the values fixed by the analytical formulae in Table~\ref{tab:damp}, where the damping factors were introduced to better reproduce our numerical results.}
\end{figure}

\begin{table}
\caption{\label{tab:damp} Analytical formulae for the elastic constants of nuclear pasta~\cite{Xia2023_PLB839-137769}, which are obtained by introducing damping factors to Eqs.~(\ref{Eq:El_MC}), (\ref{eq:El_ld_Rod}) and (\ref{eq:El_ld_Slab}). The coefficients are fixed by $\mu_{0}'\equiv \alpha n_d Z_\mathrm{droplet}^2/{R_\mathrm{W}}$ and $u\equiv{R_d}/{R_\mathrm{W}}$.}
\centering
\begin{tabular}{c|c|c|c} \hline \hline
                  &       &  $c_{11}-c_{12}$      &    $c_{44}$  \\  \hline
\multirow{2}{*}{Droplets} & BCC   &  $0.04908 \mu_{0}'$     &   $0.1705 \mu_{0}'/10^{10u^8}$     \\
                  & FCC   &  $0.03512 \mu_{0}'$     &   $0.1740 \mu_{0}'/10^{9u^8}$ \\ \hline
  Rods &  \multicolumn{2}{c|}{$c_{11}+c_{12}=3E_\mathrm{C}/10^{3u^8}$} &   $E_\mathrm{C} 10^{2.1\left(u^2-0.3\right)-3u^8}$     \\  \hline
\multicolumn{2}{c|}{Slabs}           &   \multicolumn{2}{c}{$c_{11}=6E_\mathrm{C} 10^{0.55u-10u^8-0.19}$}  \\ \hline
\end{tabular}
\end{table}

In Fig.~\ref{Fig:ElasDrop} we present the obtained elastic constants for the droplet phase in both BCC and FCC lattices as solid squares and circles, which are fixed according to Eqs.~(\ref{eq:dEBtoF}-\ref{Eq:c44_drop}). Based on the obtained structures of the droplet phase, the elastic constants can also be estimated with Eq.~(\ref{Eq:El_MC}), which means we consider the droplet phase as Coulomb crystals even above the neutron drip density. This is valid at small densities, where Eq.~(\ref{Eq:El_MC}) predicts similar values comparing with our numerical estimations~\cite{Xia2023_PLB839-137769}. Nevertheless, as illustrated in Figs.~\ref{Fig:DEcomp_dens}-\ref{Fig:DEcomp_all}, there exist slight discrepancies due to effects such as the dripped neutrons, charge screening, and small deviations from the spherical shapes of droplets~\cite{Zemlyakov2022_MNRAS518-3813}. The discrepancies become even greater at larger densities, where the elastic constants start to decrease after reaching their peaks and are not predicted by Eq.~(\ref{Eq:El_MC}). This is mainly due to the increment of neutron gas density as indicated in Figs.~\ref{Fig:Dens}, where the liquid-gas interface becomes less evident. By subtracting the background proton number density $n_p(R_\mathrm{W})$ and replacing $Z$ with $Z_\mathrm{droplet}=Z-4\pi{R_\mathrm{W}}^3 n_p(R_\mathrm{W})/3$, Eq.~(\ref{Eq:El_MC}) can be improved to better describe our numerical results. Nevertheless, if parameter Set 1 is adopted, replacing $Z$ with $Z_\mathrm{droplet}$ does not improve Eq.~(\ref{Eq:El_MC}) since $n_p(R_\mathrm{W})=0$, while the deformations induced by quadrupole electrostatic potential could also cause large discrepancies~\cite{Zemlyakov2022_MNRAS518-3813}. To improve Eq.~(\ref{Eq:El_MC}), we thus introduce damping factors as indicated in Table.~\ref{tab:damp}, which are represented by the solid and dashed curves in Fig.~\ref{Fig:ElasDrop} that well reproduce our numerical results. Meanwhile, the slope of symmetry energy plays an important role on the elastic properties, where the elastic constants of the droplet phase with $L = 41.34$ MeV are almost ten times larger than those with $L = 89.39$ MeV.

\begin{figure}
\includegraphics[width=\linewidth]{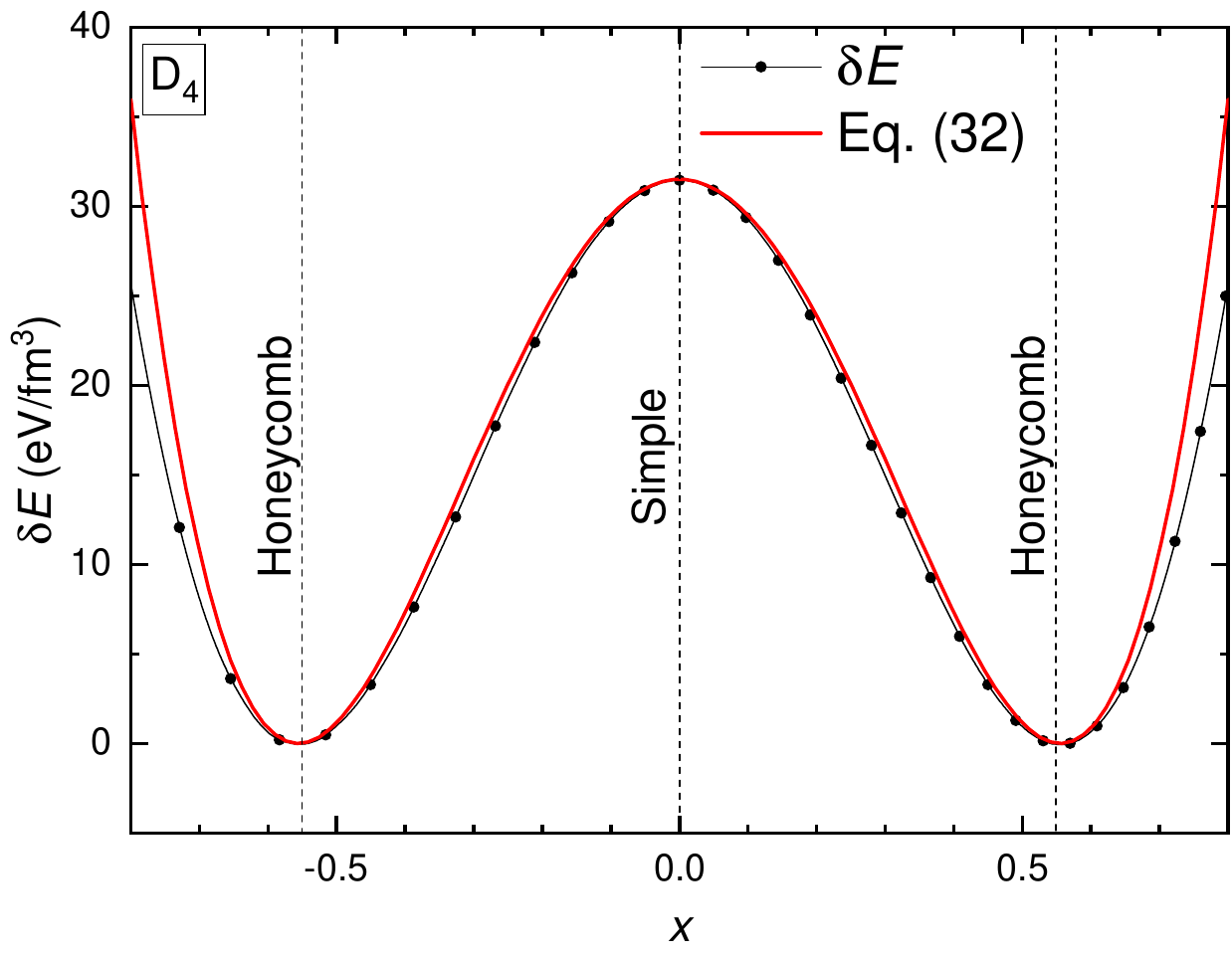}
\caption{\label{Fig:Edformc11mc22} Same as Fig.~\ref{Fig:DEBtoF} but for the rod phase at $n_\mathrm{b} = 0.06\ \rm{fm}^{-3}$ adopting parameter Set 1 with $L=41.34$ MeV.}
\end{figure}

For the rod/tube phases, by applying the deformation
\begin{equation}
    D_4:  \ \ u_{xx} = \varepsilon, \ \   u_{yy} = \frac{-\varepsilon}{1+\varepsilon}, \ \  u_{zz} = 0, \label{eq:D4}
\end{equation}
the simple configuration ($b=a$) can evolve into the honeycomb one ($b=\sqrt{3}a$) with $\varepsilon =3^{\pm1/4}-1$. In Fig.~\ref{Fig:Edformc11mc22} we present the variation of energy density with respect to deformation, where we have defined $x\equiv{\varepsilon \left( \varepsilon+2 \right) }/{(\varepsilon+1)} = (a-b)/\sqrt{ab}$. Note that $\delta E(x)$ is symmetric under reflection at $x=0$, while the simple configuration at $x=0$ corresponds to a local maximum for $\delta E(x)$, suggesting that it is unstable against decaying into the honeycomb configuration under deformation $D_4$. As indicated by the red solid curve in Fig.~\ref{Fig:Edformc11mc22}, $\delta E(x)$ can be  well described by
\begin{equation}
\delta E = \frac{3}{8} c_{44} \left[x^2-\frac{4}{\sqrt{3}}+2 \right]^2, \label{eq:dESimptoHoney}
\end{equation}
which should be valid at $|x|\lesssim 0.65$ ($-0.27\lesssim\varepsilon \lesssim 0.37$). Based on Eq.~(\ref{eq:dESimptoHoney}), we can then fix the elastic constants of the honeycomb configuration with $c_{11} - c_{12} = 2 c_{44}$. The final independent elastic constant $c_{11}$ ($=c_{22}$) is obtained adopting the deformation $D_4$ by exchanging $y$ and $z$ axis with $u_{yy}=0$, where the variation of energy density is reproduced by
\begin{equation}
  \delta E=c_{11}\varepsilon^2/2. \label{Eq:c11_rod}
\end{equation}
The elastic constant $c_{11}$ of the slab phase is determined in the same manner.

\begin{figure}
\includegraphics[width=\linewidth]{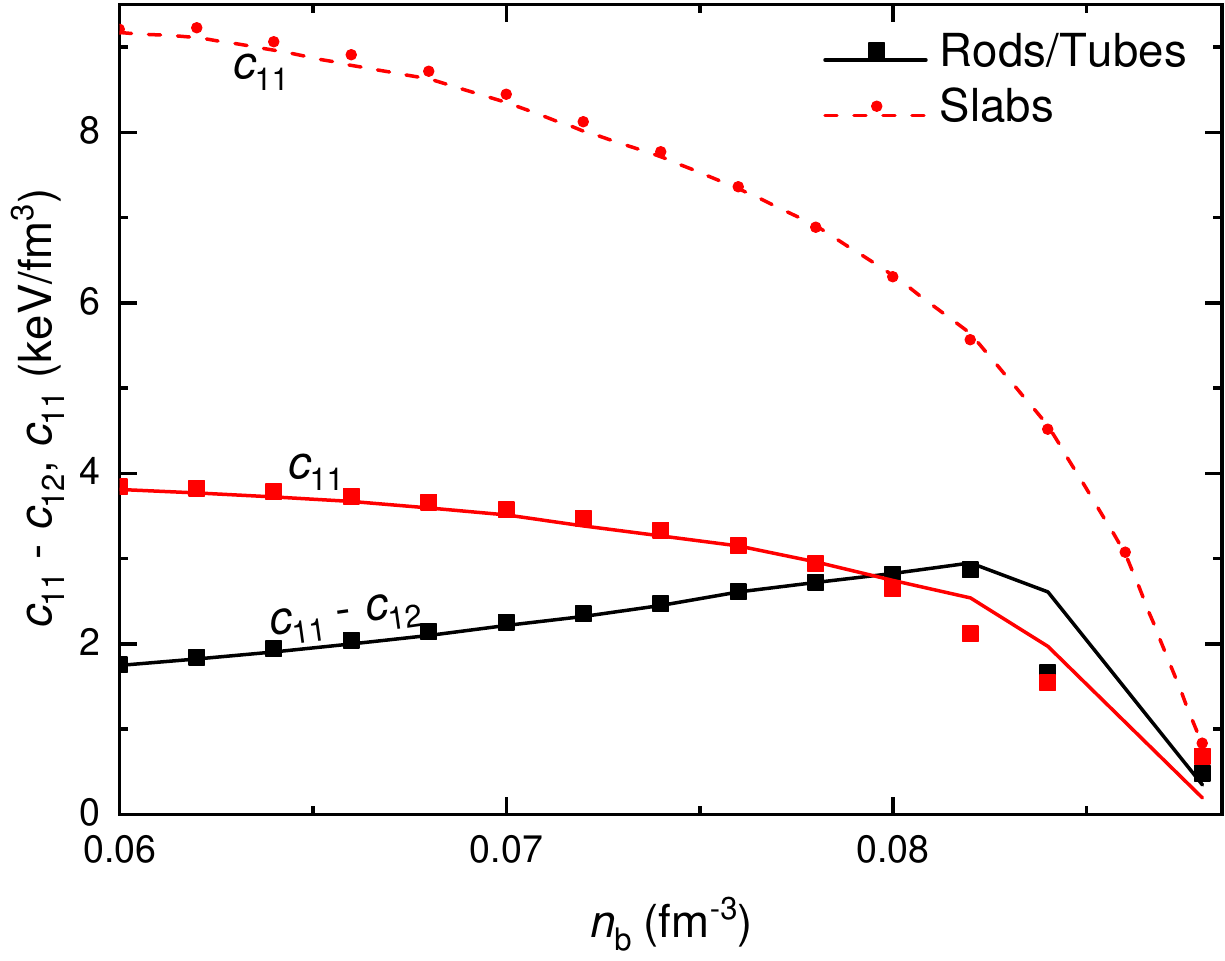}
\caption{\label{Fig:ElasPasta} Elastic constants of the rod/tube and slab phases, where the curves represent the values fixed by the analytical formulae in Table~\ref{tab:damp}. The parameter Set 1 with $L=41.34$ MeV is adopted.}
\end{figure}

The obtained elastic properties of slabs and rods/tubes in honeycomb configuration are presented in Fig.~\ref{Fig:ElasPasta}. For the rod phase in honeycomb configuration, we find that Eq.~(\ref{eq:El_ld_Rod}) gives excellent description at small densities but overestimates the elastic constants  at larger densities~\cite{Xia2023_PLB839-137769}. Meanwhile, the elastic constants of the slab phase determined by Eq.~(\ref{eq:El_ld_Slab}) deviate slightly from our estimations, where in our calculation  the thickness of a slab varies under deformation. To improve the analytical formulae in Eqs.~(\ref{eq:El_ld_Rod}) and~(\ref{eq:El_ld_Slab}), we introducing damping factors as indicated in Table~\ref{tab:damp}, which are fixed by fitting to our numerical results. The results obtained by the new formulae are indicated by the solid and dashed curves in Fig.~\ref{Fig:ElasPasta}, which predict realistic elastic constants for neutron star matter. Meanwhile, the variation of energy density under large deformations can be obtained with the polynomial expansions in Eqs.~(\ref{eq:dEBtoF}), (\ref{Eq:c44_drop}), (\ref{eq:dESimptoHoney}) and (\ref{Eq:c11_rod}).

\section{\label{sec:sum} Summary}

The elastic properties of nuclear pasta are essential to understand various astrophysical processes in neutron stars. For example, a crust destruction could trigger the sudden release of magnetic and elastic energy, causing magnetar bursts, short gamma-ray burst precursors of neutron star mergers, and pulsar glitches. The quasi-periodic oscillations observed after giant flares of soft gamma repeaters are interpreted as global oscillations of magnetars, where the oscillation spectrum is closely connected to the elastic properties of nuclear pasta. The height of mountains on the surface of neutron stars supported by the elastic stresses of astromaterials is limited by the breaking strain of neutron star crust. If the mountains are large enough causing nonzero mass quadrupoles on neutron stars, gravitational waves are emitted when they are fast rotating.

Despite the important roles played by them, various contributions from free neutrons, strong interaction among nuclei, charge screening, finite sizes of nuclei, neutron superfluidity, and weak reactions are commonly neglected in estimating the elastic properties of nuclear pasta, which lead to uncertainties when applying the results to study asteroseismology of neutron stars. In our present study, we propose a new method to estimate the elastic properties of nuclear pasta under more realistic considerations, where neutron star matter can be examined numerically in a fully three-dimensional geometry using relativistic-mean-field models with Thomas-Fermi approximation. Volume-preserving deformations are introduced to nuclear pastas with the droplets constrained to be moving along with the deforming unit cells. Based on this method, various contributions from free neutrons, strong interaction among nuclei, charge screening, finite-size effects of nuclei, and weak reactions are considered self-consistently. This enables us to obtain for the first time the realistic elastic properties of neutron star matter with a large strain
($\varepsilon \lesssim 0.5$). Our findings are listed as follows:
\begin{enumerate}
  \item The variation of energy density under large deformations are obtained, which can be reproduced by the polynomial expansions in Eqs.~(\ref{eq:dEBtoF}), (\ref{Eq:c44_drop}), (\ref{eq:dESimptoHoney}) and (\ref{Eq:c11_rod}). The elastic constants can then be fixed by comparing the $\varepsilon^2$ terms around the local minima.
  \item The obtained elastic constants agree qualitatively with those predicted by analytic formulae assuming Coulomb crystals for the droplet phase and the incompressible liquid-drop model. However, quantitative deviations are observed, which generally grow with density.
  \item By fitting to the numerical results, it is possible to improve the analytic formulae by introducing damping factors.
  \item The slope of nuclear symmetry energy plays a crucial role in the elastic properties of neutron star matter, where the the elastic constants can vary by ten times adopting $L = 41.34$ and 89.39 MeV.
\end{enumerate}

Finally, we would like to emphasize that our approach has great potential. The updated analytical formulae can readily be applied to fix the elastic properties of neutron star matter obtained with the spherical or cylindrical approximations for the Wigner-Seitz cell, which significantly reduce the numerical costs. Similarly, the elastic properties of neutron star matter at large deformations can be obtained with Eqs.~(\ref{eq:dEBtoF}), (\ref{Eq:c44_drop}), (\ref{eq:dESimptoHoney}) and (\ref{Eq:c11_rod}). The effects of temperature, superfluidity, and magnetic field can be included naturally in our future studies.

\section*{ACKNOWLEDGMENTS}
This work was supported by the National Natural Science Foundation of China(Grant Nos.~12275234 and 12342027), the National SKA Program of China (Grant No.~2020SKA0120300) and  JSPS KAKENHI (Grants No.~20K03951 and No.~20H04742).


%

\end{document}